\documentclass[modern]{aastex62}
\hypersetup{linkcolor=blue,citecolor=blue,filecolor=cyan,urlcolor=blue}

\usepackage{braket}
\usepackage{amsmath}
\usepackage{bm}
\usepackage{mathrsfs}
\usepackage{mathbbol}
\usepackage{rotating}
\usepackage{lineno}

\newcommand{\civ}{\ion{C}{4}}
\newcommand{\mgii}{\ion{Mg}{2}}

\newcommand{\het}{heteroscedasticity}

\newcommand{\hb}{H$\beta\,$}
\newcommand{\msr}{$M_\bullet-\sigma_\ast$~relation}

\revised{\today}

\shorttitle{A Generative Model for Quasar Spectra}
\shortauthors{Eilers et al.}

\sloppypar
\begin{document}

\title{\bfseries A Generative Model for Quasar Spectra}

\author[0000-0003-2895-6218]{Anna--Christina~Eilers}\thanks{NASA Hubble Fellow}
\affiliation{MIT Kavli Institute for Astrophysics and Space Research, 77 Massachusetts Ave., Cambridge, MA 02139, USA}

\author[0000-0003-2866-9403]{David~W.~Hogg}
\affiliation{Center for Cosmology and Particle Physics, Department of Physics, New York University, 726~Broadway, New York, NY 10003, USA}
\affiliation{Max Planck Institute for Astronomy, K\"onigstuhl 17, 69117 Heidelberg, Germany}
\affiliation{Center for Computational Astrophysics, Flatiron Institute, 162 Fifth Avenue, New York, NY 10010, USA}

\author[0000-0002-8177-0925]{Bernhard~Sch\"olkopf}
\affiliation{Max Planck Institute for Intelligent Systems, T\"ubingen, Germany}

\author[0000-0002-9328-5652]{Daniel~Foreman-Mackey}
\affiliation{Center for Computational Astrophysics, Flatiron Institute, 162 Fifth Avenue, New York, NY 10010, USA}

\author[0000-0003-0821-3644]{Frederick~B.~Davies}
\affiliation{Max Planck Institute for Astronomy, K\"onigstuhl 17, 69117 Heidelberg, Germany}

\author[0000-0002-4544-8242]{Jan--Torge~Schindler}
\affiliation{Max Planck Institute for Astronomy, K\"onigstuhl 17, 69117 Heidelberg, Germany}
\affiliation{Leiden Observatory, Leiden University, Niels Bohrweg 2, NL-2333 CA Leiden, Netherlands}


\correspondingauthor{Anna-Christina Eilers}
\email{eilers@mit.edu}

\begin{abstract}\noindent
We build a multi-output generative model for quasar spectra and the properties of their black hole engines, based on a Gaussian process latent-variable model.
This model treats every quasar as a vector of latent properties such that the spectrum and all physical properties of the quasar are associated with non-linear functions of those latent parameters; the Gaussian process kernel functions define priors on the function space.
Our generative model is trained with a justifiable likelihood function that allows us to treat heteroscedastic noise and missing data correctly, which is crucial for all astrophysical applications.
It can predict simultaneously unobserved spectral regions, as well as the physical properties of quasars in held-out test data. 
We apply the model to rest-frame ultraviolet and optical quasar spectra for which precise black hole masses (based on reverberation mapping measurements) are available.
Unlike reverberation-mapping studies, which require multi-epoch data, our model predicts black hole masses from single-epoch spectra, even with limited spectral coverage.
We demonstrate the capabilities of the model by predicting black hole masses and unobserved spectral regions. We find that we predict black hole masses at close to the best possible accuracy.
\end{abstract}

\keywords{\raggedright Gaussian~Processes~regression --- nonparametric inference --- astrostatistics techniques --- quasars --- spectroscopy --- supermassive~black~holes}

\section*{} 

\section{Introduction}\label{sec:intro}

Machine learning algorithms mainly fall into the two categories of \emph{supervised} and \emph{unsupervised}.
In supervised learning tasks, data points that consist of features and labels are used to train the model, such that it can classify data into different categories (when the labels are discrete), or model a continuous relationship between features and labels (when the labels are real-valued or lists of real values).
Unsupervised learning algorithms generally aim to understand the structure of a data set, uncover patterns in the data or cluster unlabeled data sets.
An equally important distinction in machine learning is between \emph{discriminative} models and \emph{generative} models.
Discriminative models are designed to find functions of data features that predict data labels.
Generative models are designed to find functions that predict features, or predict features given labels, or predict both features and labels.

In the natural sciences---and astrophysics in particular---generative models have an advantage over discriminative models, which is that they can naturally include any peculiar uncertainties, noise, and missing-data properties in the data set. This is possible because the generative model synthesizes the features; it can be trained with a justifiable loss function such as a log-likelihood that contains a reasonable representation of the noise model in the feature space, and drops missing data.
This is in contrast to discriminative models, most of which require complete, rectangular data, which are implicitly believed to be correct or true.

One approach to generative, unsupervised machine learning is to represent a complex high dimensional data set in a lower-dimensional latent space. The archetypal such method is principal component analysis (PCA). PCA seeks a linear projection of the data onto a lower-dimensional subspace, represented by an orthonormal basis, such that the data generated from the low-dimensional PCA preserves as much variance as possible from the original data.
Because the input data can be reconstructed from the PCA projections, there is a sense in which PCA can be thought of as a generative model. Since PCA is linear, it is often not capable of representing the full structure in the data through its linear low-dimensional embedding.

A Gaussian process latent variable model (GPLVM) is a fully probabilistic, non-linear, generative latent-variable model that generalizes PCA \citep{Lawrence2003, Lawrence2005, LawrenceMoore2007, TitsiasLawrence2010}. It is a generative model that represents a non-linear extension of the linear probabilistic PCA \citep{TippingBishop1999}, and has been shown to be a powerful approach for probabilistic modelling of high dimensional data through dimensionality reduction. The classical GPLVM is unsupervised; it does not distinguish features and labels and is not designed for supervised learning tasks. 

In this work, we aim to introduce and provide a framework to apply the GPLVM to astrophysical settings. In particular, we are interested in problems that come with noisy, heteroscedastic, complex data sets containing both spectral features and labels, where some parts of the data set might be missing or unobserved. To this end, we construct a modified version of the standard GPLVM and expand the algorithm to a ``multi-output'' generative model \citep[see also][]{Gao2011}. This multi-output GPLVM generates \textit{both} the features as well as their associated labels simultaneously from a common latent space (see \S~\ref{sec:model}), and thus enables predictions of both the spectral features as well as the labels. 

As a first application (see \S~\ref{sec:application}), we apply this generative model to quasar spectra aiming to determine their physical properties, such as the masses of their central accreting supermassive black holes (SMBHs), based on their single-epoch spectra alone. Determining accurate black hole masses for quasars is challenging and usually requires time-intensive, multi-epoch observations with regular cadence to conduct reverberation mapping (RM) measurements \citep[e.g.][]{Peterson1993, Barth2015, Shen2016}, which are currently unfeasible for quasars beyond redshift $z\gtrsim 2$. Thus, for the vast majority of quasars black hole mass estimates are obtained by means of scaling relations which relate the quasar's luminosity and the emission line widths observed in their spectra to black hole mass estimates calibrated based on low-redshift reverberation mapped quasars \citep[e.g.][]{VestergaardPeterson2006, Grier2017, Coatman2017}. However, we know that information about the quasars' black hole masses is encoded in the single-epoch quasar spectra, which we aim to reveal by means of our generative model. 

In the chosen generative model we present here, both the quasar spectra as well as their physical labels (such as their black hole masses) are simultaneously generated from points in a latent space.
Because the data set contains missing (or unobserved) data (and because the non-missing data have heteroscedastic noise properties), the generative model has the advantage that every extant piece of data can be handled appropriately, and missing features and labels are no problem. The generative model can also correctly account for different data measurement precisions (i.e.\ data weights). 

The single biggest issue with the model is the small size of the available training data. There are currently only 31 quasars that meet our data-quality cuts. Despite this, the model makes good predictions for black hole masses and spectral pixels in held-out data. We will discuss this and other limitations of the model in \S~\ref{sec:limitations}, and highlight possible future applications and improvements in \S~\ref{sec:summary}.

\section{A Gaussian Process Latent Variable Model}\label{sec:model}

The GPLVM is a generative model that represents a flexible, non-linear approach for a dimensionality reduction using a Gaussian process to learn a low-dimensional representation of (potentially) high-dimensional data. Typically, the GPLVM is used for completely non-supervised learning tasks, but here we modify the algorithm to a multi-output generative model, such that it can generate \emph{both} data features as well as its associated labels simultaneously. Furthermore, the GPLVM can naturally handle data sets with heteroscedastic uncertainties as well as missing or unobserved data, which is crucial for any applications to real astronomical problems. However, dealing with heteroscedastic noise and missing data implies that the implementation of the GPLVM becomes significantly more complicated and less scalable to large data sets, which we discuss further in \S~\ref{sec:limitations}. 

\subsection{Assumptions}

Our model makes a number of strong assumptions: We assume that our data is correct, in the sense that the measurements are unbiased, with normally distributed noise, and do not include substantial outliers. Relatedly, we assume that all measurements, i.e. spectral pixels as well as labels, are independent measurements 
with no (or negligible) covariances (although we will later weaken this assumption in Eqn.~\ref{eq:Kx}). We assume that the high-dimensional data set in the joint space of labels and features (spectral pixels) is drawn from a distribution that is intrinsically low dimensional in nature, such that each data point can be represented as a point in a lower-dimensional latent space. Additionally -- and importantly -- we assume that the mapping of the latent space to observations
can be expressed as draws from a Gaussian process.

\subsection{Input data}

Let us assume we have $N$ training-set objects. Their features (e.g. quasar spectral pixels) make up a $N\times D$ rectangular matrix $X$ (this matrix will however, have many missing values, see \S~\ref{sec:missing_data}), which are given as 
\begin{equation}
    X = [\vec{x}_1, ..., \vec{x}_N]^T ~~\text{with uncertainties}~~ \sigma_X = [\vec{\sigma}_{x_1}, ..., \vec{\sigma}_{x_N}]^T ~,
\end{equation}
where the $\vec{x}_n$ are individual $D$-vector spectra.
Associated with these features is an $N\times L$ rectangular matrix $Y$ of labels
\begin{equation}
    Y = [\vec{y}_1, ..., \vec{y}_N]^T ~~\text{with uncertainties}~~ \sigma_Y = [\vec{\sigma}_{y_1}, ..., \vec{\sigma}_{y_N}]^T ~,
\end{equation}
where the $\vec{y}_n$ are individual $L$-vector labels. In our example application below (see \S~\ref{sec:application}) the labels include the black hole mass, bolometric luminosity and redshift of the quasars, but other physical properties can easily be added as additional labels. 
Technically, the uncertainty information could be full covariance matrices, but in this case we will treat the uncertainties as independent, such that the uncertainties can be represented with objects the same sizes as $X$ and $Y$.

For a stable optimization of the model we re-scale the input data set, such that each of the $D$ columns of $X$ and the $L$ columns of $Y$ have zero mean and unit variance. The $\vec{x}_n$ and $\vec{\sigma}_{x_n}$ are scaled consistently, as are the $\vec{y}_n$ and the $\vec{\sigma}_{y_n}$. 

\subsection{Handling missing data}\label{sec:missing_data}

Although the input data $X$ and $Y$ are technically rectangular, i.e. 2-dimensional matrices of shape $N\times D$ and $N\times L$, respectively, where every record has the same length and the same feature structure, in practice there is a lot of missing data. For instance, some labels might not be measured for a subset of objects, or the spectra in the data set might be observed with different telescopes or instruments resulting in a different wavelength coverage. Additionally, we will transform all quasars to their rest-frame wavelengths (see \S~\ref{sec:training_data}), and thus even quasars observed with the same telescope and instrument but at slightly different redshifts will also have a different rest-frame wavelength coverage.These missing data entries yield to sparse data matrices $X$ and $Y$. 

The standard GPLVM can handle the missing data in a conceptually rigorous way by only taking objects into account which have finite data, i.e. measured values, at any given pixel $d$ or label $l$. In this way, also objects with missing or unknown labels $l$ can be accounted for in the training set. The missing data and uncertainties in the rectangular input data are represented with NaNs. 

\subsection{Kernel functions}
The idea of the GPLVM is that the state of object $n$ can be represented with a $Q$-dimensional latent vector $\vec{z}_n$. These can be combined into a rectangular $N\times Q$ latent variable block $Z$ 
\begin{equation}
    Z = [\vec{z}_1, ..., \vec{z}_N]^T. 
\end{equation}
This one set of latents will generate all the spectra and all the labels, or in other words the $Q$-vector $\vec{z}_n$ will generate the spectrum $\vec{x}_n$ and labels $\vec{y}_n$ of object $n$.

The Gaussian processes are defined by kernel functions; the kernel functions determine the prior over functions. For the kernel function relating the features $X$ to the latents we choose the commonly used radial basis function (RBF) kernel 
\begin{equation}
    K_x(z_i, z_j) = A_x\exp\left[-\frac{B_x}{2}\,(z_i-z_j)^T(z_i-z_j)\right], \label{eq:Ktheta}
\end{equation}
while we choose different hyperparameters but the same kernel function relating the labels $Y$ to the latents, i.e.\ 
\begin{equation}
    K_{y}(z_i, z_j) = A_{y, l}\exp\left[-\frac{B_{y}}{2}\,(z_i-z_j)^T(z_i-z_j)\right].   \label{eq:Kgamma}
\end{equation}
The hyperparameters $A_x$ and $A_{y, l}$ constitute the amplitude of the kernels, while the hyperparameters $B_x$ and $B_{y}$ denote the length scales of the kernels. Note there is a redundancy between the length scales and the overall scale of the latent variables $Z$, which is ameliorated by adding a prior on the latent parameters (third term in Eqn.~\ref{eq:log_post}). However, in practice we find that 
we obtain a better optimization of the model when keeping the scale lengths $B_x$ and $B_{y}$ simply fixed to unity. Note that we take only one kernel function for all $D$-dimensional features (Eqn.~\ref{eq:Ktheta}), while we use a different kernel for each label $l$ (Eqn.~\ref{eq:Kgamma}), i.e.\ the hyperparameter $A_{y, l}$ can be different for each label $l$. In what follows we will refer to the set of hyperparameters as $\theta = [A_x, A_{y, l=0}, ..., A_{y, l=L}]$.

\subsection{Accounting for heteroscedasticity in the input data}
The quasar spectra $X$ as well as the labels $Y$ have measurement uncertainties $\sigma_X$ and $\sigma_Y$, respectively, which can naturally be accounted for in a GPLVM. In order to account for the \het\ in the data, we construct covariance matrices for each spectral pixel $d$ and each label $l$, i.e.
\begin{equation}
    C_d = \begin{bmatrix} \sigma^2_{x_{1,d}} & 0 & ... & ...\\ 0&\sigma^2_{x_{2,d}}&0& ... \\  \vdots&\vdots&\ddots&\vdots\\... & ... & 0 &  \sigma^2_{x_{N,d}}\end{bmatrix}
    ~~\text{and}~~ 
    C_l = \begin{bmatrix} \sigma^2_{y_{1,l}} & 0 & ... & ...\\ 0&\sigma^2_{y_{2,l}}&0& ... \\  \vdots&\vdots&\ddots&\vdots\\... & ... & 0 &  \sigma^2_{y_{N,l}}\end{bmatrix}
\end{equation}
that will be added to the kernel functions, i.e.\ 
\begin{align}
    \hat{K}_{x, d} &= K_x + (1+\beta)\, C_d \label{eq:Kx} \\ 
    \hat{K}_{y, l} &= K_y + C_l. 
\end{align}
The term $(1+\beta)$ has been added to capture the ``information content'' of the spectra. If we completely trusted the noise model in our quasar spectra we would set $\beta=0$, while a higher value of $\beta$ indicates that the noise in the quasar spectra might be underestimated. Thus in practice $\beta$ represents another hyperparameter to our model. Note that this represents a tractable solution for capturing the covariant structure of the spectra, but a much better and more correct approach would be to use a Gaussian process in the spectral dimension. However, given the structure of our model, this would be extremely difficult to make computationally tractable and is thus beyond the scope of this paper. 

The kernel matrices $\hat{K}_{x, d}$ and $\hat{K}_{y, l}$ have dimensions $N_d\times N_d$ and $N_l\times N_l$, respectively, where $N_d$ and $N_l$ denote the number of objects in the training set that have a measured and finite data point at pixel $d$ or label $l$. 

\subsection{Likelihood functions}

The logarithm of the joint probability of the features $X$, the labels $Y$, and the latent variables $Z$ in the GPLVM represents the ``cost function'' and is given as 
\begin{equation}
    \ln\mathscr{L}(X, Y | \theta, Z) = \ln\mathscr{L}_x(X|\theta, Z) + \ln\mathscr{L}_y(Y|\theta, Z) + \ln\mathscr{P}_z(Z).  \label{eq:log_post}
\end{equation}
The first term in Eqn.~\ref{eq:log_post} denotes the likelihood function for the input features $X$, which includes a sum over all pixels, assuming that all pixels can be treated independently. The likelihood function is conditioned on the latent variables $Z$ and the hyperparameters $\theta$, i.e.
\begin{equation}
    \ln\mathscr{L}_x(X|\theta,Z) = \sum_{d=1}^D\left[-\frac{N_d}{2}\ln2\pi-\frac{1}{2}\ln(\det\hat{K}_{x, d}) - \frac{1}{2}\left(X^T_{:, d}\,\hat{K}^{-1}_{x, d}\,X_{:, d}\right)\right], \label{eq:Lx}
\end{equation}
where $X_{:, d}$ denotes the input features of all quasars at pixel $d$, and $N_d$ denotes the number of objects with given finite input data at pixel $d$. 

Analogously, the second term in Eqn.~\ref{eq:log_post} describes the likelihood function for the labels $Y$, conditioned on the latent variables $Z$ and the hyperparameters $\theta$, i.e.\  
\begin{equation}
    \ln\mathscr{L}_y(Y|\theta,Z) = \sum_{l=1}^L\left[-\frac{N_l}{2}\ln2\pi-\frac{1}{2}\ln(\det\hat{K}_{y, l}) - \frac{1}{2}\left(Y^T_{:, l}\,\hat{K}^{-1}_{y, l}\,Y_{:, l}\right)\right].  \label{eq:Ly}
\end{equation}
Here, $Y_{:, l}$ indicate the label values of label $l$ for all objects, while $N_l$ denotes the number of objects with given label $l$.

The third term in Eqn.~\ref{eq:log_post} constitutes a \textit{prior} on the latent variables $Z$, for which we choose a Gaussian with zero mean and unit variance, i.e.\ 
\begin{equation}
    \ln\mathscr{P}_z = -\frac{N}{2}\ln2\pi - \frac{1}{2}Z^2. \label{eq:Lz}
\end{equation}
As mentioned previously, this prior is useful in alleviating the redundancy between the latent variables and the scale lengths $B_x$ and $B_y$. Nevertheless, there are still many exact degeneracies between the different latent dimensions, in the sense that this model is rotationally symmetric in the latent space $Z$. 

Note that Eqn.~\ref{eq:Lx} and \ref{eq:Ly} follow the same mathematical structure and thus could be combined to one likelihood function treating the labels as a second set of features. Similarly, one could split Eqn.~\ref{eq:Ly} into multiple separate likelihood functions for each individual label $l$, while mathematically, the structure of our model would still remain unchanged. We consider this GPLVM, where the latent space generates multiple different outputs -- in this case the spectral features of quasars as well as their physical labels -- a ``multi-output'' generative model. 

In contrast to standard Gaussian process regression where one fits for the data, our model optimizes for the latent parameters $Z$ as well as the hyperparameters $A_x$ and $A_{y, l}$ that maximize the cost function in Eqn.~\ref{eq:log_post} for the data. 
Because the $X$- and $Y$-space predictions are non-linear functions of the latents $Z$, the optimizations are non-linear and not guaranteed to find a global optimum.
Note that technically this model is not fully Bayesian, because at training time parameters are found by optimization.

\section{Model predictions}\label{sec:testing_step}

\subsection{``$z$-step'': finding a set of latent parameters for new input data}\label{sec:zstep}

Once we have optimized the GPLVM with respect to the latent parameters $Z$ as well as the set of hyperparameters $\theta$, we can apply it to a new and yet unseen data point with spectrum $x_\ast$\footnote{Note that we omit the vector notation here for better readability, but please keep in mind that throughout this manuscript the testing objects naturally have the same dimensions as the training set data, i.e.\ $x_\ast$ is a $D$-dimensional vector and $y_\ast$ an $L$-dimensional vector.}, which could have a subset of already known labels $y_\ast$, in order to predict its set of unknown labels $\tilde{y}_\ast$ and possibly missing or unobserved spectral pixels $\tilde{x}_\ast$. We first find (by optimization) the set of latent parameters $z_\ast$ that represents the new input $\{x_\ast, y_\ast\}$, which will then be used to estimate the unknown part of the data. 

A benefit of Gaussian processes \citep[e.g.][]{Williams1998, Lawrence2005, williams2006gaussian} is that any new data point $\{x_\ast, y_\ast\}$ (with measurement uncertainties $\{\sigma_{x_\ast}, \sigma_{y_\ast}\}$) that is not part of the training set will have a posterior probability distribution function (PDF) that is a Gaussian. That is,
\begin{equation}
    \mathscr{L}(x_\ast, y_\ast|z_\ast) = \mathcal{N}\left(x_\ast, y_\ast | \mu_{x} (z_\ast), s_{x}^2(z_\ast), \mu_{y} (z_\ast), s_{y}^2(z_\ast)\right),  \label{eq:like_for_z}
\end{equation}
with means $\mu_{x}$ and $\mu_{y}$, and variances $s_{x}^2$ and $s_{y}^2$ (given in Eqn.~\ref{eq:mux} to \ref{eq:sigmay} below). 

The set of latent variables $z_\ast$ for the new data point $\{x_\ast, y_\ast\}$ is determined by computing the likelihood of the observed data given the projection of the posterior probability estimate for $z_\ast$ back into the data-space, i.e.\ 
\begin{align}
    \ln\mathscr{L}(x_\ast, y_\ast|z_\ast, X, Y, Z, \theta) = &-\frac{1}{2}\ln(s_x^2 + \sigma^2_{x_\ast}) -\frac{1}{2}\,\frac{\left(x_{\ast}-\mu_x\right)^2}{s_x^2 + \sigma^2_{x_\ast}} - \frac{N_d}{2}\ln(2\pi) \nonumber \\
    &-\frac{1}{2}\ln(s_y^2 + \sigma^2_{y_\ast}) -\frac{1}{2}\,\frac{\left(y_{\ast}-\mu_y\right)^2}{s_y^2 + \sigma^2_{y_\ast}} - \frac{N_l}{2}\ln(2\pi) \label{eq:test_step_likelihood}
\end{align}
with a mean for the feature space $X$
\begin{equation}
    \mu_{x} = X^T \hat{K}^{-1}_x k_x(Z, z_\ast)\label{eq:mux}
\end{equation}
and variance
\begin{equation}
    s_{x}^2 = k_x(z_\ast, z_\ast) - k_x(Z, z_\ast)^T\hat{K}_x^{-1}k_x(Z, z_\ast),  \label{eq:sigmax}
\end{equation}
and analogously for the label space $Y$, i.e.\ 
\begin{align}
    \mu_y = Y^T \hat{K}_y^{-1}k_y(Z, z_\ast)\label{eq:muy}
\end{align}
and 
\begin{align}
    s_y^2 &= k_y(z_\ast, z_\ast) - k_y(Z, z_\ast)^T \hat{K}_y^{-1}k_y(Z, z_\ast).   \label{eq:sigmay}
\end{align}
Here, $k_x(Z, z_\ast)$ and $k_y(Z, z_\ast)$ denote the column vector constructed from computing the elements of the kernel matrices between the training set and the new point in latent space $z_\ast$ \citep[see e.g.][for details]{Williams1998, Lawrence2005, williams2006gaussian}. 

Note that the posterior distribution over $Z$ could be multi-modal with respect to $z_\ast$. Thus one should use sampling methods to evaluate the posterior distribution, and then approximate $z_\ast$ around the largest mode. However, in practice, we do not find any multi-modality for the chosen application (see \S~\ref{sec:application}), and hence simply take the set of parameters $z_\ast$ which maximizes Eqn.~\ref{eq:test_step_likelihood}. 

\subsection{Predicting unknown labels $\tilde{y}_\ast$ and missing spectral pixels $\tilde{x}_\ast$}

Once we have a set of latent variables $z_\ast$ that represent the new data point $\{x_\ast, y_\ast\}$, the prediction for the \textit{unknown} labels $\tilde{y}_\ast$ also follows a Gaussian distribution \citep[e.g.][]{Lawrence2005}, i.e.\  
\begin{equation}
    \mathscr{L}(\tilde{y}_\ast|z_\ast, Y, Z, \theta) = \mathcal{N}\left(\tilde{y}_\ast|\mu_y(z_\ast, Y, Z, \theta), s_y^2(z_\ast, Z, \theta)\right) \label{eq:predict_y_given_z}
\end{equation}
with a mean and variance given in Eqn.~\ref{eq:muy} and \ref{eq:sigmay}. Note that this step does not require any further optimization. 

Given the set of latent variables $z_\ast$ that represents the new quasar in the latent space, we can not only predict its unknown labels but also its missing spectral pixels $\tilde{x}_\ast$, as the latent variables in our model represent \textit{both} the features, i.e.\ spectra, as well as the labels. Note that technically this is exactly the same as predicting the unknown labels $\tilde{y}_\ast$ in Eqn.~\ref{eq:predict_y_given_z}, and hence each (missing) spectral pixel of an object, which can either be part of the training set or a new unseen spectrum, is given by
\begin{equation}
    \mathscr{L}\left(\tilde{x}_\ast|z_\ast, X, Z, \theta\right) = \mathcal{N}\left(\tilde{x}_\ast | \mu_{x} (z_\ast, X, Z, \theta), s_{x}^2(z_\ast, X, Z, \theta)\right), \label{eq:predict_x_given_z}
\end{equation}
where $\mu_x$ and $s_x^2$ are given by Eqn.~\ref{eq:mux} and \ref{eq:sigmax}.

\subsection{Predicting spectral features for given labels: sampling the latent space}

Just as it is possible to predict missing pixels from the feature vector, it is possible to predict the entire feature vector given an input label. However, the problem with this is that the assumption that the data are uniquely represented as a point in the latent $Z$-space becomes untrue if the only input to the test-step likelihood is a single label and no spectrum. Thus, in order to understand how quasar spectra depend on their labels, such as the black hole mass for instance, we sample the $Q$-dimensional latent space and find regions that correspond to sets of latent parameters representing quasars with the given input label. 

In practice, we take Gaussian random draws from the $Q$-dimensional latent space and determine the corresponding label $y_\ast$ using Eqn.~\ref{eq:predict_y_given_z}. We then find regions in the latent $Z$ space which represent the same label $y_\ast$ and determine their spectral features using Eqn.~\ref{eq:predict_x_given_z}. By averaging the spectral features from many latent representations of the same label $y_\ast$ we can search for spectral dependencies on the labels. As a first example, we will show later how quasar spectra depend on their physical properties (see Fig.~\ref{fig:dependencies} in \S~\ref{sec:application}). 

\section{Implementation notes}\label{sec:implementation}

We use the L-BFGS-B algorithm \citep{Zhu1997} to optimize the cost function in Eqn.~\ref{eq:log_post} of our model. We optimize simultaneously the $Q$-dimensional latent parameters $Z$ for each object in our training set, as well as for the hyperparameters describing the amplitudes of the kernel functions $A_x$ and $A_{y, l}$, where the latter can be different for each label $l$. Thus, in practice we are optimizing $N \cdot Q + (L+1)$ parameters. We take the derivatives analytically and include the Jacobian matrix for better optimization. 

The latent dimension $Q$ as well as the hyperparameter $\beta$ from Eqn.~\ref{eq:Kx} are optimized via cross-validation. Since our main goal of a first application described in \S~\ref{sec:application} will be to predict the black hole masses from single-epoch quasar spectra, we train the GPLVM $N$ times on $N-1$ quasars, and afterwards predict the black hole mass label of the omitted $N$th object. This procedure is then repeated for different values for $Q$ and $\beta$. We chose a set of parameters $Q$ and $\beta$ that minimizes the bias and scatter in the distribution of predicted black hole masses compared to the input black hole masses in this cross-validation (see Fig.~\ref{fig:1to1} in \S~\ref{sec:application}). 

Our model permits (and requires!) many choices regarding not only the optimization criteria for hyperparameters, but also the input and output parameters. Furthermore the choice of the kernel functions (Eqn.~\ref{eq:Ktheta} and \ref{eq:Kgamma}) is a strong model assumption and could be optimized by cross-validation. The choices of the model parameters and kernel functions will depend on the specific goals and applications of the GPLVM. In the next section when we apply our model to quasar spectra, we will make different choices with regards to the number $L$ and chosen labels when predicting the quasars' black hole masses (see Fig.~\ref{fig:1to1}) or when understanding the spectral dependencies of quasar spectra on various physical properties (see Fig.~\ref{fig:dependencies}). Note that we do not conduct an extensive parameter search to find the ``best'' model choices for our application, and thus the model might perform better with a different, more optimized set of parameters for a given application and science goal. 

We will discuss the limitations of our current model implementation as well as advanced algorithms that will significantly improve the performance of our GPLVM in the future in detail in \S~\ref{sec:limitations}. 

\section{A first application: Predicting physical properties of quasars from their spectra}\label{sec:application}

Our multi-output generative GPLVM has numerous potential applications. Here we want to provide a first example, where we apply the model to single-epoch quasar spectra in order to predict black hole mass measurements from their spectral features alone. We briefly summarize how SMBHs are measured in quasars to give some context (\S~\ref{sec:context}), before introducing our data set of quasar spectra for which the masses of their central supermassive black holes (SMBHs) are well known (\S~\ref{sec:training_data}). 
We will show that the prediction accuracy of the SMBH masses from the GPLVM is as good as the measurements allow (\S~\ref{sec:results_MBH}) and further show how our model can predict missing or unobserved spectral pixels (\S~\ref{sec:results_spectra}). At the end we demonstrate how the quasar spectra depend on their physical parameters  (\S~\ref{sec:results_pca}).

\subsection{Context: Measuring the masses of supermassive black holes}\label{sec:context}

For nearby galaxies the masses of their central SMBHs can be measured by resolving the sphere of influence around the black hole in the motion of stars or gas, which has resulted in the now well established $M_\bullet-\sigma_\ast$ relation at $z\sim 0$ \citep[e.g.][]{Magorrian1998, Gebhardt2000, HaeringRix2004, Gultekin2009} relating the mass of a black hole $M_\bullet$ in the center of galaxies to the stellar velocity dispersion $\sigma_\ast$ in the galactic bulge. For quasars, this is not feasible, since the central accreting black hole outshines the stellar light by several orders of magnitudes. 

For a few quasars at low redshifts, i.e.\ $z\lesssim1$, precise mass estimates have been obtained via the so-called \textit{reverberation mapping} (RM) technique or \textit{echo mapping} \citep[e.g.][]{BlandfordMcKee1982, Peterson1993}, which enables measurements of the distance between the SMBH in the center of a quasar and gas clouds orbiting the black hole within the so-called broad-line region (BLR). Assuming that the gas motion in the BLR is completely dominated by the gravitational pull of the black hole, one can then derive the black hole mass using Newton's law of motion, i.e.
\begin{align}
    M_{\bullet} &= f \frac{R_{\rm BLR}\, \Delta v^2}{G} = f\,\frac{c\,\tau\,\sigma_{\rm rms}^2}{G}, \label{eq:rm_bh}
\end{align}
where $\Delta v$ denotes the width of the emission lines broadened by Doppler broadening due to the velocity of the orbiting gas clouds, which can be inferred from the variance $\sigma_{\rm rms}^2$ of the emission lines in the spectra of the quasars, $G$ is the gravitational constant, and $f$ is a geometric factor to account for the unknown geometry and gas distribution of the BLR. The radius of the BLR, $R_{\rm BLR}$, is estimated by means of the RM method, which measures the time lag $\tau$ between changes in the continuum emission arising from the accretion disk around the black hole and the corresponding line emission changes from the gas clouds, once the radiation has propagated outwards from the black hole to the broad-line region. Thus the radius of the broad line region can be estimated as $R_{\rm BLR} = c\tau$, with $c$ as the speed of light. 

The average geometric factor $f$ is determined for an ensemble of objects by enforcing that the black hole mass measurements fall on the well-known local $M_\bullet-\sigma_\ast$ relation \citep[e.g.][]{Onken2004}, which has, however, an intrinsic scatter of $0.35-0.5~\rm dex$ \citep[e.g.][]{Magorrian1998, McLureDunlop2002}, and therefore represents the limiting precision for all black hole mass measurements. Additional uncertainties can arise, if there is a redshift evolution of the $M_\bullet-\sigma$ relation \citep[e.g.][]{Pensabene2020}, or if quasars were to obey a different scaling relation than quiescent galaxies without nuclear activity \citep[e.g.][]{Woo2015}. 

Since the RM measurements require long monitoring and expensive observations, for most quasars -- especially at higher redshifts where time delays are longer due to time dilation and the generally more massive black holes -- this method is unfeasible. Thus, masses of SMBHs for most quasars are commonly inferred from single-epoch spectra by using scaling relations that relate the width of an emission line and the quasar's luminosity to the black hole mass \citep[e.g.][]{VestergaardPeterson2006, Grier2017, Coatman2017}. These scaling relations are calibrated based on quasars with precise RM black hole mass measurements. However, additional uncertainties arise for quasars at high redshifts beyond $z\gtrsim 3$ for two reasons: First, the high-redshift quasars have generally more massive black holes and are more luminous than the observed quasar sample at lower redshift, requiring an extrapolation of the scaling relations in a parameter space which is only sparsely sampled at lower redshifts. Second, since the rest-frame optical emission lines, such as \hb, commonly used for calibrating the scaling relations cannot be measured with ground-based observatories at high redshifts anymore, additional scaling relations between the width of the \hb\ emission line and rest-frame UV lines, such as \ion{Mg}{2} or \ion{C}{4}, are required to estimate the black hole masses \citep[e.g.][]{Wang2009}. These various scaling relations and extrapolations make black hole mass estimates for most quasars in the universe highly uncertain and potentially biased. 

With the generative model presented here we aim to circumvent all scaling relations and intend to constrain the black hole masses of quasars (and other physical properties) directly from the single-epoch spectra themselves. 

\subsection{Training data: quasar spectra with precise black hole mass measurements}\label{sec:training_data}

Our training data set consists of $31$ quasar spectra for which reliable RM black hole mass measurements have been reported in the literature \citep[e.g.][]{Bentz2009, Barth2015}. We chose only quasars with reliable \hb\ emission line time lags due to the large scatter in the measured time lags observed between different emission lines \citep[e.g.][]{Grier2017, Fausnaugh2017}. 

The quasar spectra are obtained with two different instruments, i.e.\ the Space Telescope Imaging Spectrograph (STIS) and the Cosmic Origins Spectrograph (COS) on the Hubble Space Telescope (HST), covering the rest-frame UV and optical wavelengths \citep{Park2013, Park2017}. We transform all spectra to rest-frame wavelengths and take all observed spectral pixels between $1220-5000$~{\AA} into account. This wavelength range is chosen to avoid any absorption from the intergalactic medium bluewards of the Ly$\alpha$ emission, and to include the \hb\ emission line at $\lambda_{\rm rest}\approx4861$~{\AA}. Due to the different redshifts of the quasars, i.e.\ $0.002\leq z\leq 0.234$, all quasars cover slightly different rest-frame wavelengths. Thus, the data set is highly heteroscedastic with a lot of missing data and also varying data quality, i.e.\ with signal-to-noise ratios between $5\lesssim{\rm S/N}\lesssim 107$. 

In order to prepare the rectangular input data we apply a cubic spline fit to each quasar spectrum and afterwards apply a $3\sigma$-clipping to mask any absorption lines within the quasar continua, which arise due to intervening foreground absorption systems along the quasar sightline and are thus not intrinsic to the quasars themselves. We then fit a power-law continuum, i.e.\ $f_\lambda\propto\lambda^{-\alpha}$ to the spectral regions free of emission lines in the quasar spectra, and re-scale all spectra to be approximately unity at $\lambda_{\rm rest}=2500$~{\AA} by dividing the spectra by the value of the power-law continuum at this wavelength. We bin all spectra to a common wavelength grid between $1220~{\rm \AA} \leq \lambda \leq 5000~{\rm \AA}$ with a fixed pixel scale of $\Delta\lambda=2$~{\AA} without correlating the noise of neighbouring pixels. All pixels that are unobserved or have been masked are set to NaNs. Furthermore, for a more stable optimization we pivot and scale all pixels to have a mean of zero and unity variance. This results in a rectangular input data set  $X$ of shape $N\times D$. The matrix $\sigma_X$ contains the corresponding measurement uncertainties on each pixel value. 

For this first example, we choose either two or three labels for each quasar: we always take the quasars' black hole masses $M_\bullet$ and bolometric luminosities $L_{\rm bol}$ as labels, as well as for some applications we also take the quasars' redshifts $z$ into account (e.g.\ Fig.~\ref{fig:dependencies}). However, the label vector could easily be augmented by additional quasar properties, such as for example the Eddington ratio of their mass accretion rate $\lambda_{\rm Edd}$. For our chosen data set all labels are measured, but in principle, any missing or unknown labels could be set to NaNs. Thus, we could include objects in the training set for which we do not have (or do not wish to include) certain label measurements. In the end we construct two rectangular matrices $Y$ and $\sigma_Y$ of shape $N\times L$ containing the input labels and their uncertainties, respectively. 

The uncertainties on the redshift $z$ arise solely from measurement uncertainties of the peak of the \hb\ emission line used to derive the quasars' redshifts. Uncertainties on $L_{\rm bol}$ also only contain measurement uncertainties of the monochromatic luminosity $\lambda L_{1350}$, which we transform into a bolometric luminosity using the bolometric correction factor of $4.3$ \citep{Richards2006, VestergaardOsmer2009}. The uncertainty on the black hole mass estimates arise from a combination of factors: First, we have measurement uncertainties in the time lag $\tau$ as well as the line width $\sigma_{\rm rms}$. Second, the dominating uncertainty in the black hole mass measurements arises from the geometric (or virial) factor $f$, which relates the measured virial product to the black hole mass estimates by calibrating the measurements to the local \msr\ \citep[e.g.][]{Onken2004, Woo2015}. We use a recent measurement by \citet{Woo2015} for the virial factor, i.e.\ $\log_{10} f=0.65\pm 0.12$, which the authors derive by jointly fitting the \msr\ using local quiescent galaxies and reverberation-mapped active galactic nuclei. Using this virial factor we update the black hole mass measurements for our data sample reported in \citet{Park2013, Park2017}. 

Note that the virial factor $f$ can only be determined for an ensemble of galaxies and quasars, by requiring that the black hole mass estimates fall \textit{on average} onto the local \msr. Thus, all black hole mass measurements -- even when precisely determined via RM measurements -- have an intrinsic scatter of approximately $0.35-0.5$~dex \citep[e.g.][]{VestergaardPeterson2006, VestergaardOsmer2009, Woo2015, Park2017}, which constitutes a systematic uncertainty on all black hole mass measurements. This intrinsic scatter limits the precision in the black hole mass estimates we can possibly achieve. 

\begin{figure}
    \centering
    \includegraphics[width=0.9\textwidth]{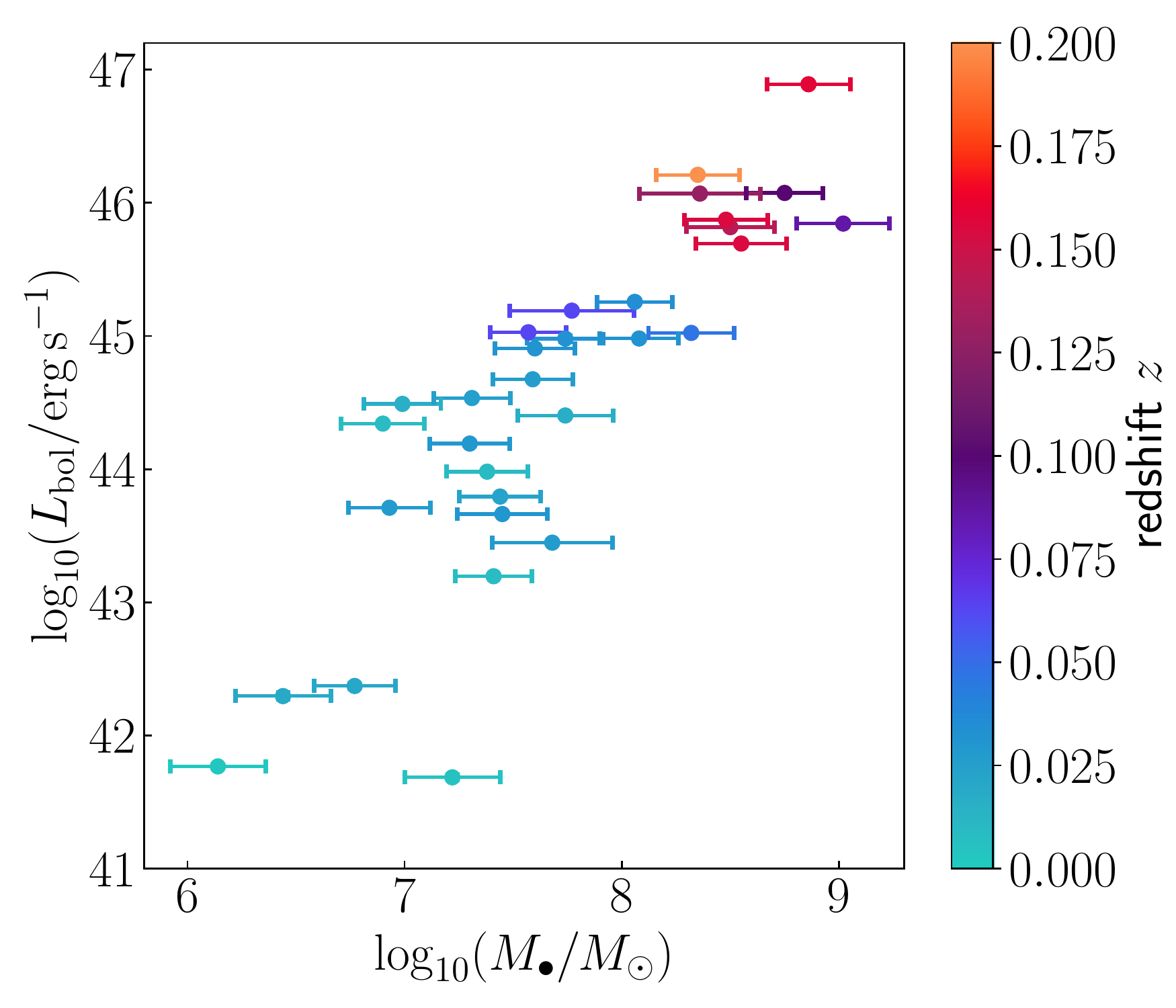}
    \caption{Black hole masses and bolometric luminosities colored by redshift of 31 quasars in our data sample, which have reliable black hole mass measurements based on the RM technique. }
    \label{fig:parent_sample}
\end{figure}

\begin{sidewaysfigure}
    \centering
    \includegraphics[width=0.49\textwidth]{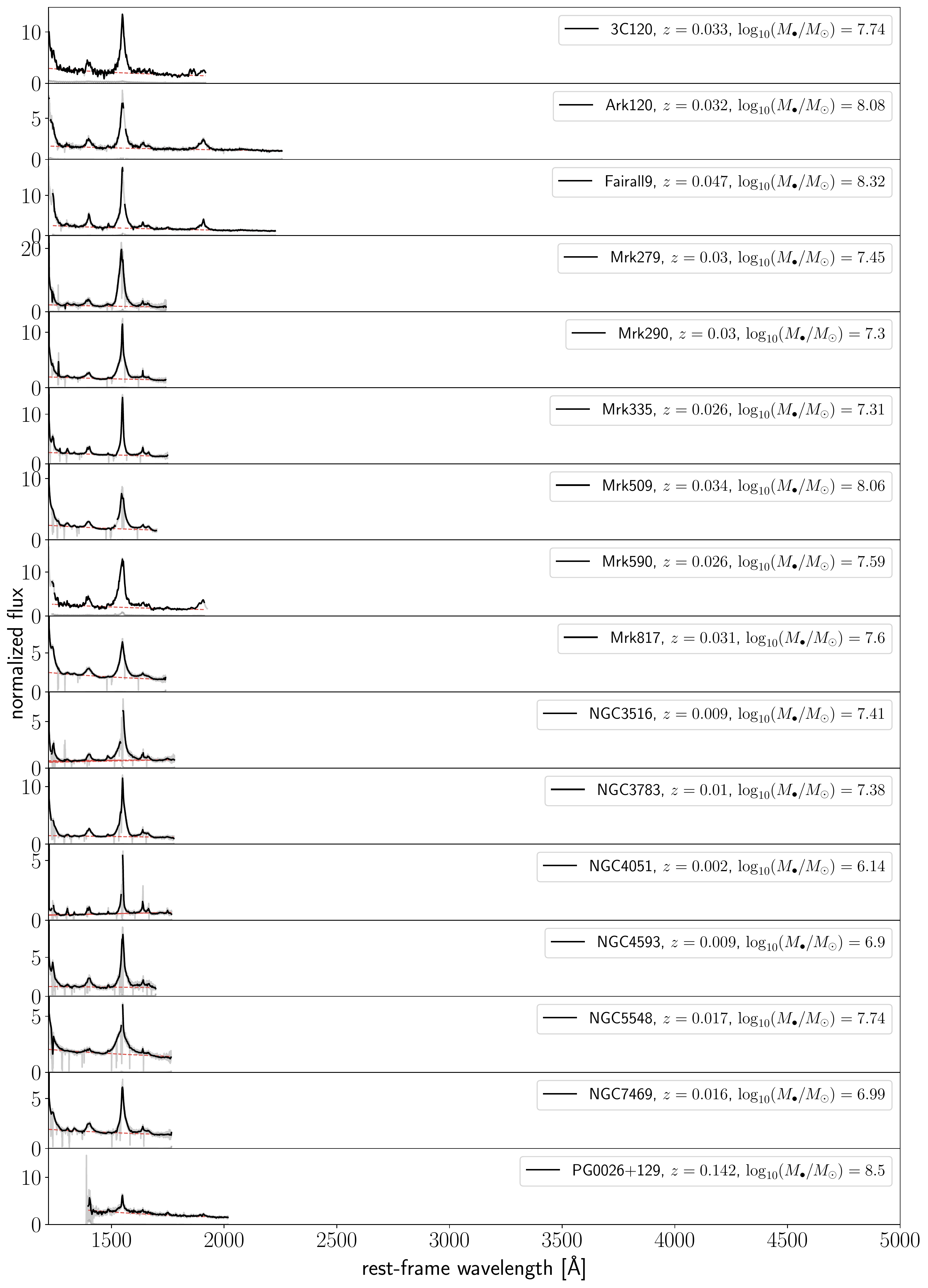}
    \includegraphics[width=0.49\textwidth]{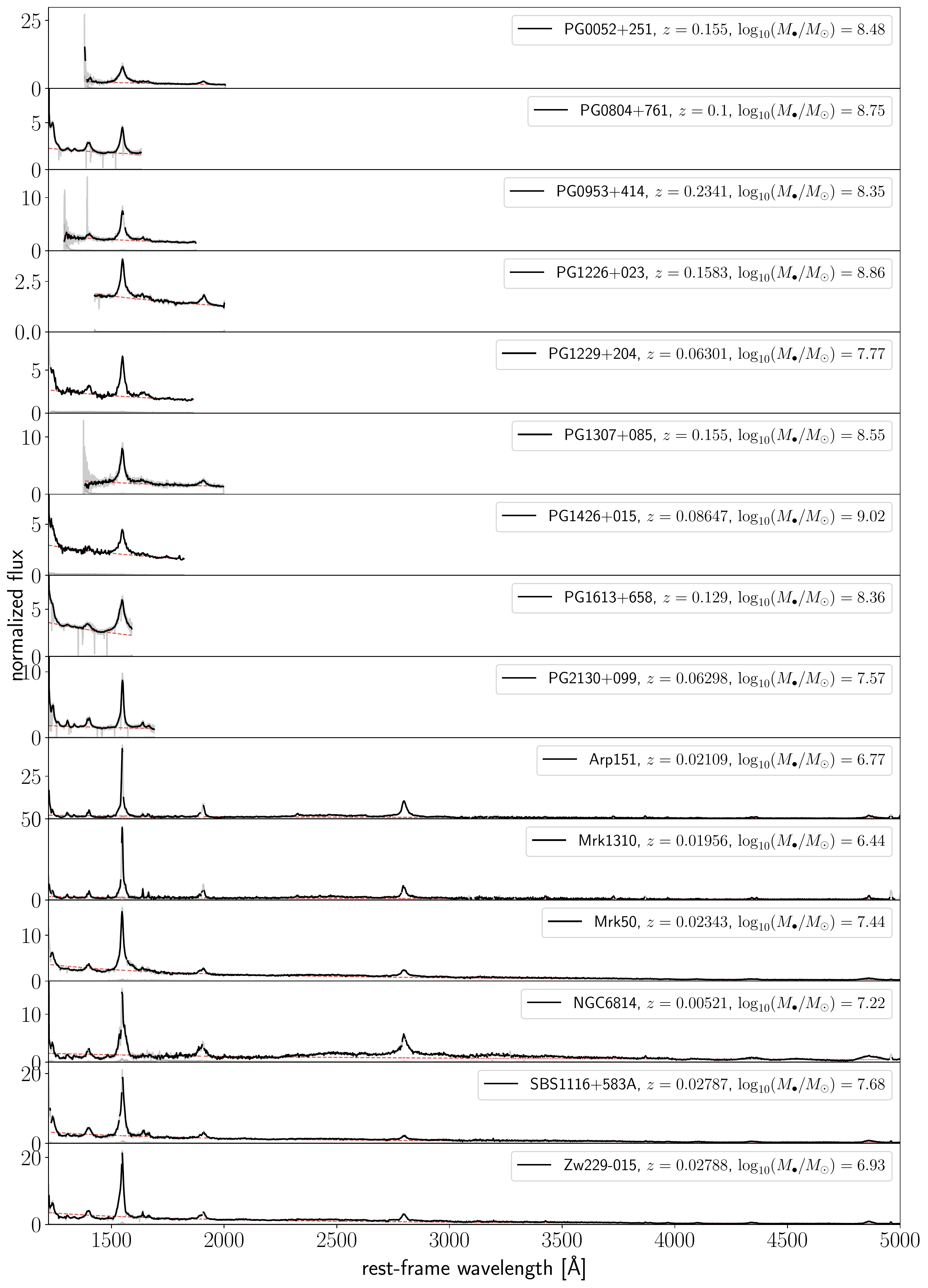}
    \caption{Quasar spectra in our data set taken with COS/HST and STIS/HST. The light grey spectra show the original data, while the black curves show the input data re-binned to a common wavelength grid and after fitting a spline and sigma-clipping in order to remove absorption lines. The red dashed curve shows the power-law fit to each spectrum, which is used to normalize all spectra to unity at approximately $2500$~{\AA}. }
    \label{fig:spectra}
\end{sidewaysfigure}

The properties of the quasars in our data set are shown in Fig.~\ref{fig:parent_sample} and the spectra are shown in Fig.~\ref{fig:spectra}\footnote{Note that we exclude one object (3C390) from the data set presented in \citet{Park2017} due to its very unusual emission line shape, caused either by a strong \ion{N}{4}] $\lambda1486${\AA} emission line \citep{Park2013} on top of the \ion{C}{4} $\lambda1549${\AA} emission line or alternatively a strong foreground absorption feature. }. 

\subsection{Predicting black hole masses from single-epoch quasar spectra}\label{sec:results_MBH}

\begin{figure}
    \centering
    \includegraphics[width=0.9\textwidth]{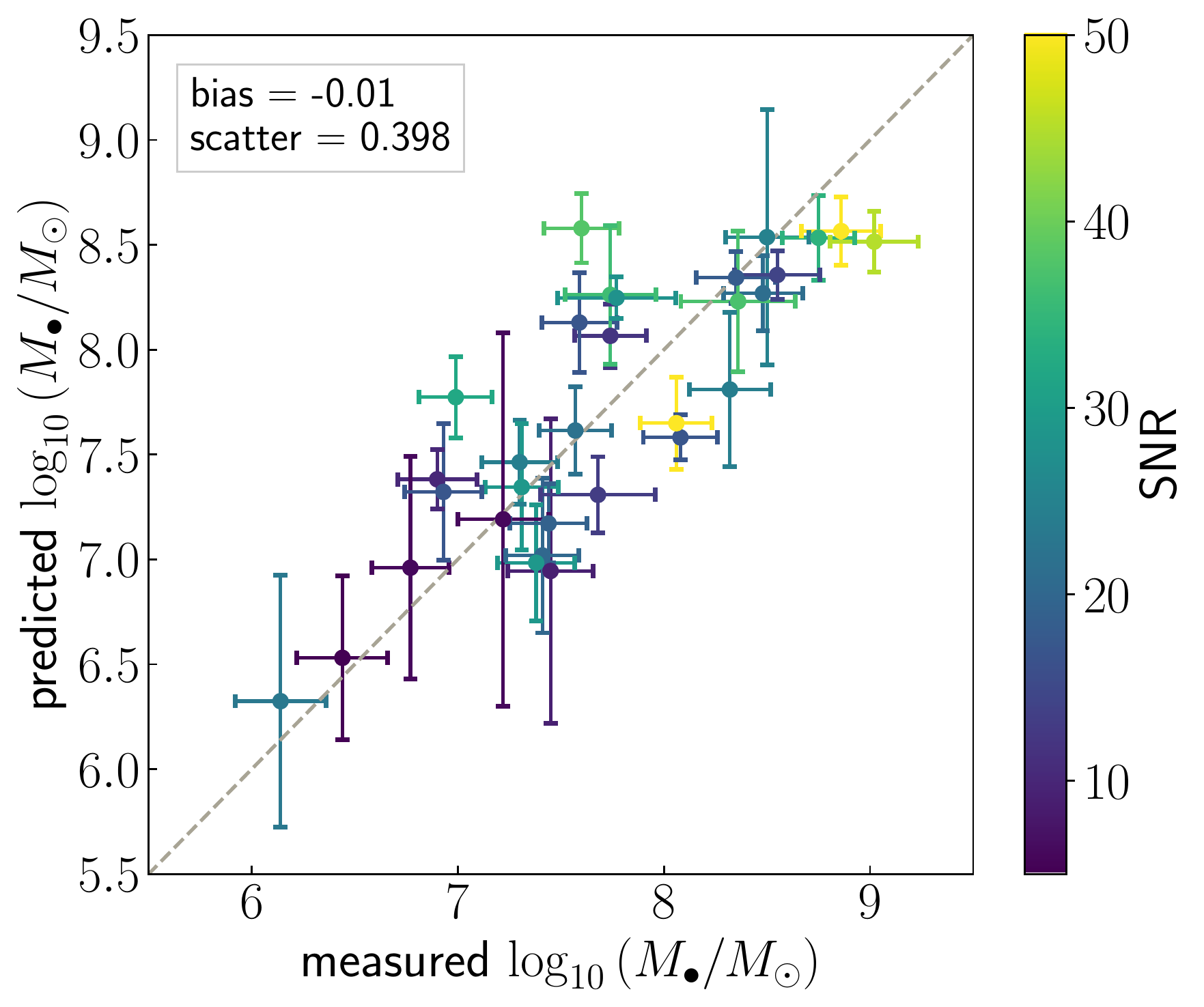}
    \caption{Cross-validation of the GPLVM with $Q=16$ latent dimensions. Predicted compared to measured black hole masses for all quasars in our data set, colored by the SNR of the single-epoch quasar spectrum. Note that we use only two labels for the quasars here, i.e.\ $L=2$, namely black hole mass and bolometric luminosity. }
    \label{fig:1to1}
\end{figure}

As described in \S~\ref{sec:implementation} we train the GPLVM with different parameters for the latent dimension $Q$ and the hyperparameter $\beta$ and determine the best values for these parameters by cross-validation of the predicted black hole masses compared to the input black hole mass labels. We find the best predictions for the black hole masses for values of $Q=16$ and $\beta=10$, shown in Fig.~\ref{fig:1to1}. The predicted black hole mass labels are nearly unbiased with an offset of $\sim1\%$ and have a scatter of approximately $0.4$~dex, which is en par with the best possible precision we can achieve, since it agrees with the intrinsic scatter of the RM measured black hole masses \citep[e.g.][]{VestergaardPeterson2006, Woo2015}. 

While the uncertainties on the measured input black hole masses are all comparable in size, since the uncertainty on the virial factor $f$ dominates the error budget, the errorbars on the predicted black hole masses reflect the ``information content'' of the input data, i.e.\ the sizes of the errorbars correlate somewhat with the signal-to-noise ratio of the spectrum, as well as with the spectral coverage of the input spectrum. 

\begin{figure}
    \centering
    \includegraphics[width=\textwidth]{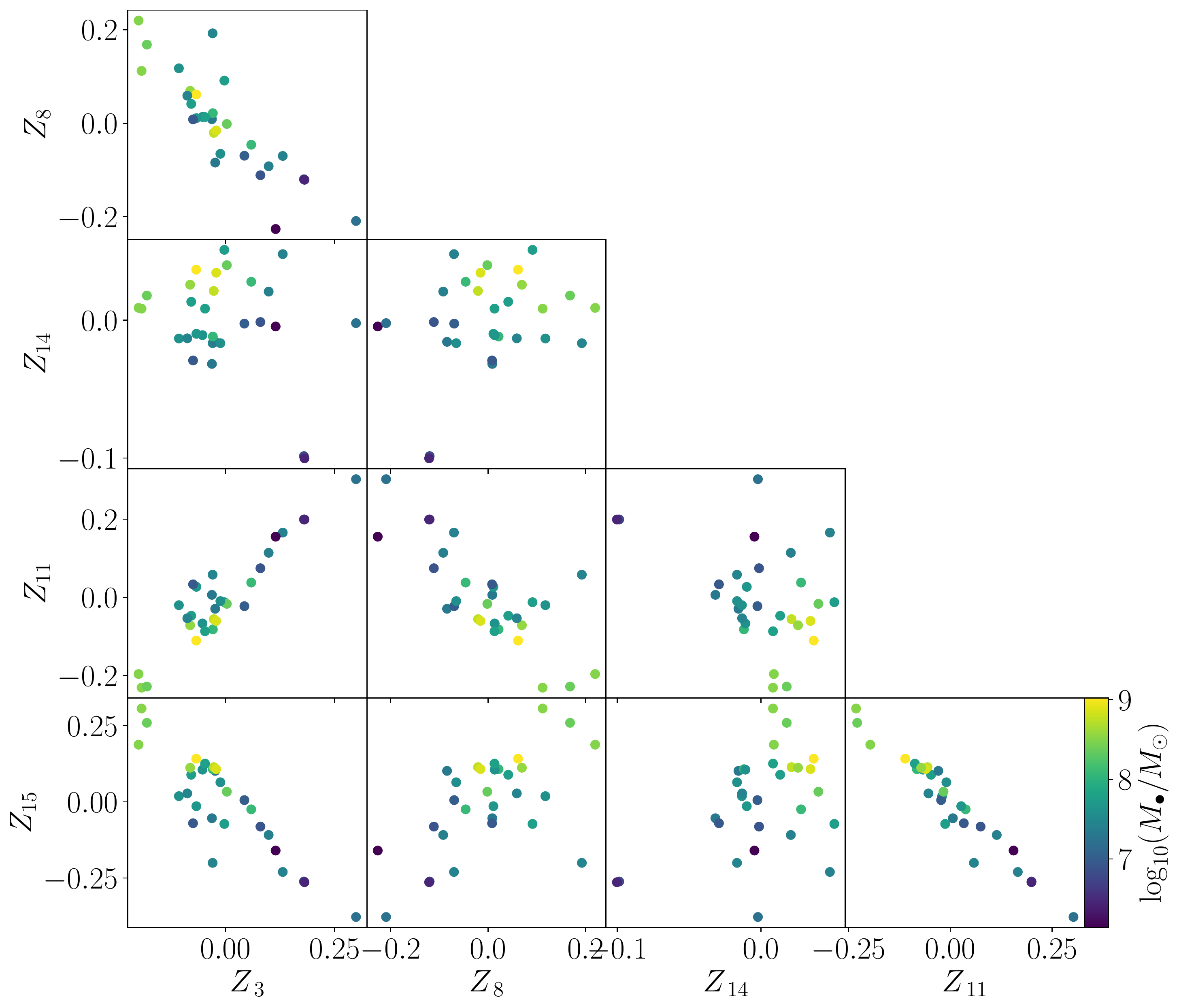}
    \caption{Visualization of the latent space. We show the locations of the quasars in five latent dimensions, which show the strongest gradient with black hole mass based on the Pearson correlation coefficient. }
    \label{fig:corner}
\end{figure}

In Fig.~\ref{fig:corner} we visualize four dimensions of the latent space, which show the strongest dependency on the black hole mass label as determined from calculating the Pearson correlation coefficients. It is evident that not one latent dimension alone encodes the information of the black hole mass, but rather the black hole masses depend on a combination of the $16$ latent dimensions. Note that the latent dimensions are strongly degenerate and hence by modifying the prior on the latent space (Eqn.~\ref{eq:Lz}), one could in principle impose a dependency of the black hole mass onto a chosen latent dimension. In practice we find that imposing a more restrictive prior does not improve the predictions. 

It is important to note that in the regime where all quasars have complete spectral coverage of certain emission lines (i.e. in the here chosen data set all quasars have spectral coverage of the \civ\ emission line), the scatter of $0.4$~dex in the predicted black hole masses shown in Fig.~\ref{fig:1to1} is comparable to the scatter in predictions from scaling relations \citep[e.g.][]{Park2017}. However, these scaling relation are no longer applicable for objects where these emission lines are unobserved, masked by telluric absorption or simply very noisy, in which case the GPLVM can still produce a reliable black hole mass estimate. Furthermore, the GPLVM allows us to include more quasars in the training step that might not have coverage of certain spectral features, but would nevertheless improve modeling (see \S~\ref{sec:summary}). 

\subsection{Predicting unobserved or ``missing'' spectral regions}\label{sec:results_spectra}

\begin{figure}
    \centering
    \includegraphics[width=\textwidth]{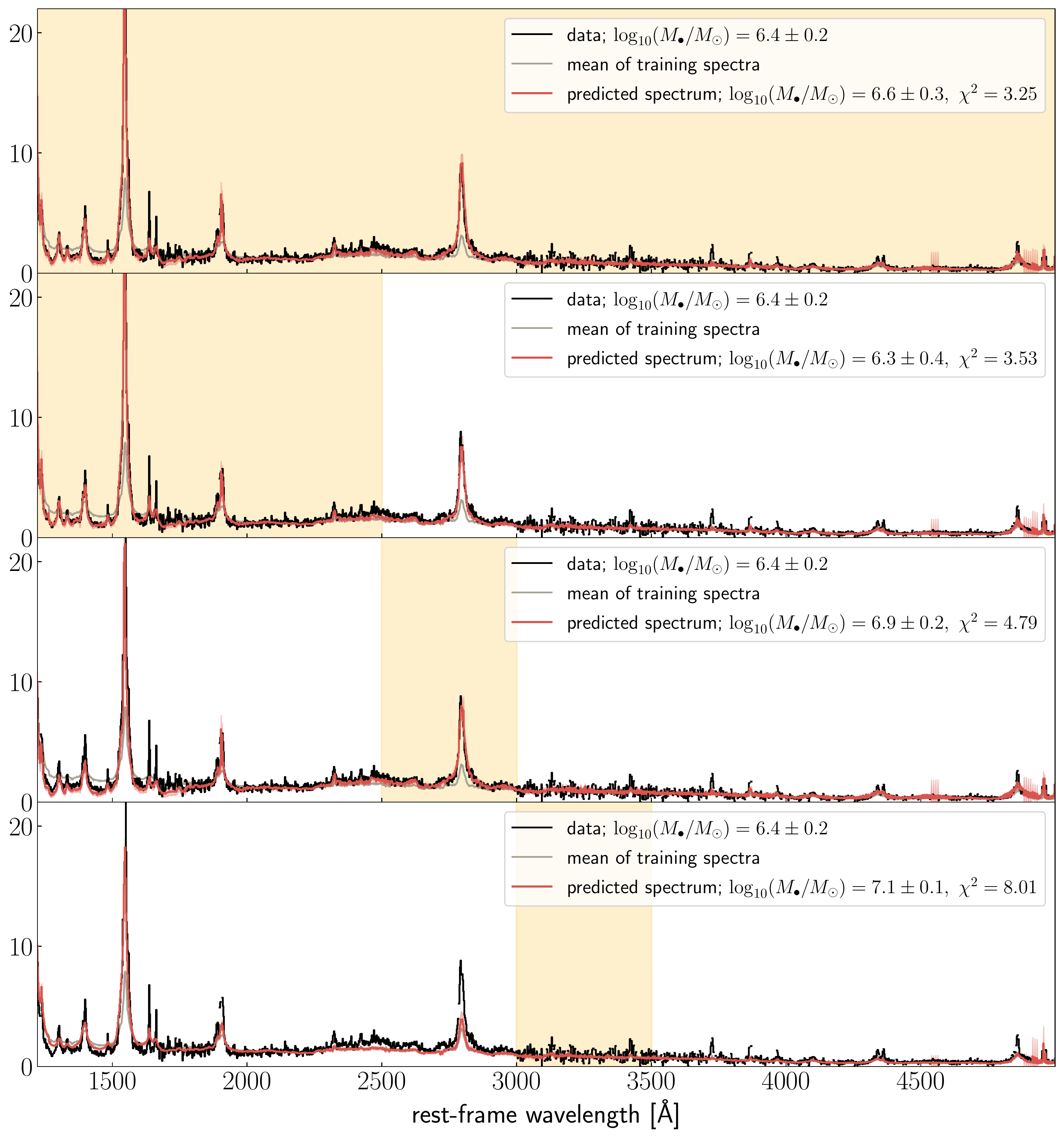}
    \caption{Predictions of missing spectral regions. In each panel we use the spectral region indicated by the yellow shaded regions to determine the latent representation of the new unseen quasar spectrum shown in black (see \S~\ref{sec:zstep}). The red curves show the predicted spectra derived from the set of latent parameters. As expected the precision of the spectral prediction increases when a larger fraction of the spectrum is used to determine the latent parameters, whereas the prediction approaches the mean of the training set spectra shown in grey, if only a small fraction of the spectral coverage is provided as input. }
    \label{fig:pred_spectra}
\end{figure}

In Fig.~\ref{fig:pred_spectra} we test how well the spectral features of a new quasar spectrum that was not part of the training set can be predicted by our generative model. In each panel we use the yellow-shaded spectral region $x_\ast$ (without any information about the labels $y_\ast$) to determine the latent parameters $z_\ast$ using Eqn.~\ref{eq:test_step_likelihood}. With this set of latent parameters we generate the ``missing'' spectral regions $\tilde{x}_\ast$ and labels $\tilde{y}_\ast$ by means of Eqn.~\ref{eq:predict_y_given_z} and \ref{eq:predict_x_given_z}. 

Generally, the predictions of held-out spectral features works extremely well as long as the spectral coverage of the input features contain sufficient information. 
However, the accuracy of the spectral feature prediction as well as the black hole mass label prediction decreases with less spectral coverage as expected, and approaches something like the mean of the training-set spectra shown in grey, if the input $x_\ast$ is not sufficiently informative.  
With only 30 quasars in each leave-one-out training set, it is perhaps surprising that these predictions are so good.
The quality of these predictions indicates that quasar spectra are intrinsically very low in dimensionality, since even limited spectral coverage and a limited data set can train a model that makes good predictions (with a reduced $\chi^2\lesssim 5$, see Fig.~\ref{fig:pred_spectra}) of held-out spectral features. 

Note that the hyperparameters $Q$ and $\beta$ were determined via cross-validation when optimizing the black hole mass predictions (Fig.~\ref{fig:1to1}) rather than the spectral features, and thus it is likely that the predictions for the unobserved spectral regions could be further improved by choosing a different set of values for $Q$ and particularly for $\beta$ that would be optimized with respect to the spectral predictions. Furthermore, due to the very small number of objects in the training set the model is trained on a limited range of quasar spectra, and thus larger numbers of quasars in the training set will also improve this prediction. 

\subsection{Spectral dependencies on quasar properties}\label{sec:results_pca}

\begin{figure}
    \centering
    \includegraphics[width=\textwidth]{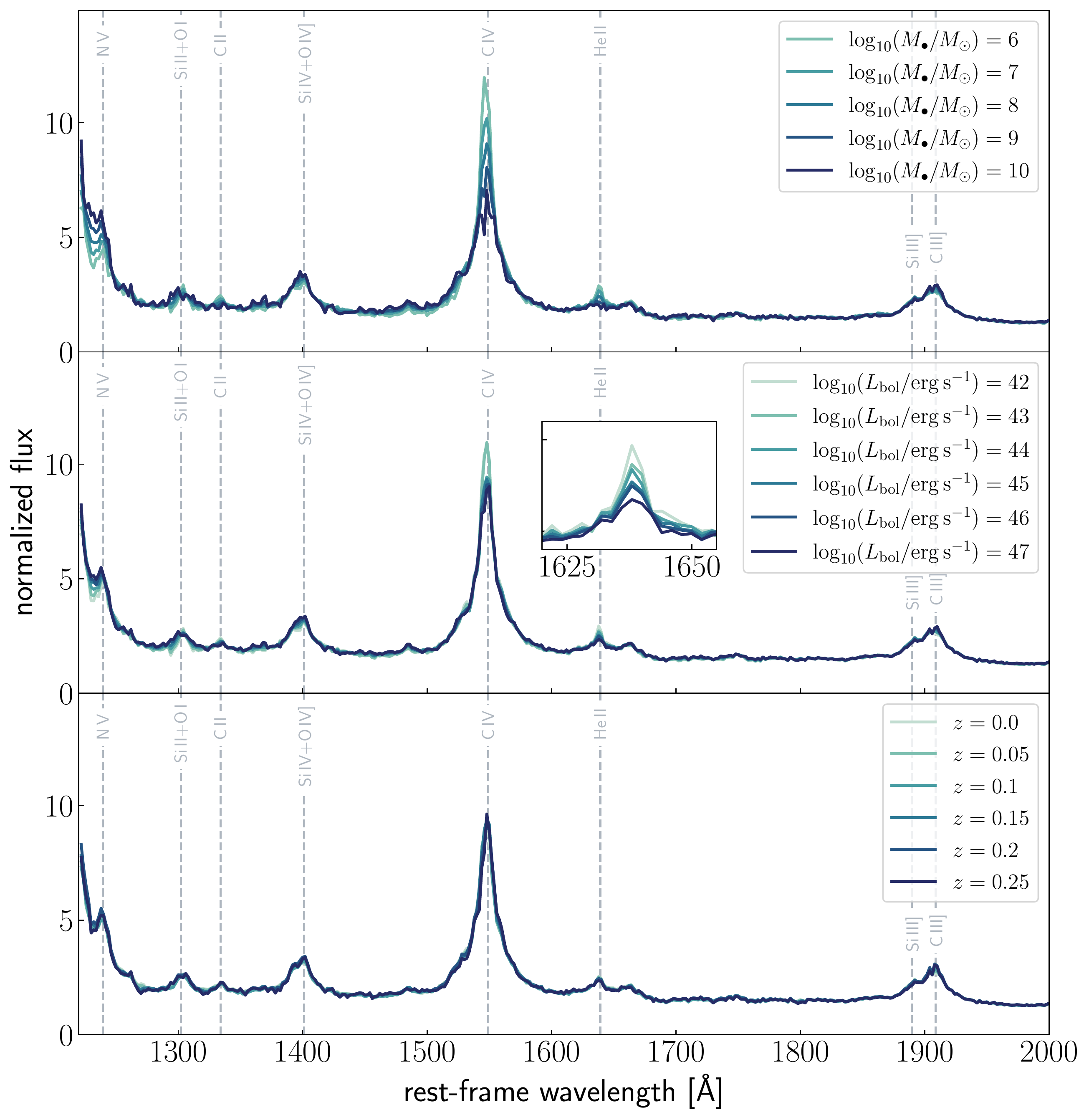}
    \caption{Spectral dependencies on various quasar properties, i.e.\ black hole mass (\textit{top}), bolometric luminosity (\textit{middle}) and redshift (\textit{bottom}). The inset panel shows the expected spectral changes due to the Baldwin effect. }
    \label{fig:dependencies}
\end{figure}

In Fig.~\ref{fig:dependencies} we show the spectral dependencies of the quasar spectra on three labels, i.e.\ black hole mass, bolometric luminosity, and redshift. We randomly sample the $16$-dimensional latent space and show the median of all spectra that correspond to a given black hole mass within $\Delta\log_{10}(M_\bullet/M_\odot)= 0.1$, a fixed bolometric luminosity within $\Delta \log_{10}(L_{\rm bol}/\rm erg\,s^{-1}) = 0.2$ and a fixed redshift within $\Delta z= 0.01$. We only show the wavelength region at $\lambda_{\rm rest}\lesssim 2000$~{\AA}, where we see the strongest spectral differences, likely due to the fact that only six quasars in our current data sample cover the rest-frame optical wavelength regime (see Fig.~\ref{fig:spectra}). 

We observe a few interesting and expected trends: In the top panel of the figure showing the spectral dependencies with black hole mass, the emission lines show the expected broadening with increasing black hole mass, such as the \ion{C}{4} emission line for instance, the \ion{S}{2}+\ion{O}{1} complex or the \ion{N}{5} emission line.  
Interestingly, the amplitude of the \ion{C}{4} emission line shows a stronger dependency on the black hole mass than the width of the line, suggesting that line width of the \ion{C}{4} emission line alone is likely not a good proxy for black hole mass \citep[see also][]{Coatman2017}. 

Furthermore, the semi-forbidden lines \ion{S}{3}] and \ion{C}{3}] show a less strong dependency on the black hole mass than the permitted lines. These lines generally arise from lower density gas close to the critical density, i.e. $N_e\sim 3\times10^9\rm\,cm^{-3}$ \citep[e.g.][]{Osterbrock2006}, which is likely located at larger radii from the black hole \citep[e.g.][]{AGN1990}, and thus the Doppler broadening is less apparent. 

The second panel shows the trends with bolometric luminosity, where we can nicely observe the Baldwin effect \citep{Baldwin1977}, which indicates that quasar spectra show a decreasing equivalent width of their UV and optical emission lines with increasing bolometric luminosity. Both the \ion{C}{4} emission line as well as some of the fainter lines such as \ion{He}{2}, which is highlighted in the inset plot, show this effect.  

We note, however, that the interpretation of this figure should be taken with caution, due to the limited number of objects that is currently used for training the GPLVM. Due to the nature of our current data set there is also a mild degeneracy between $L_{\rm bol}$ and $M_\bullet$ (see Fig.~\ref{fig:parent_sample}) and thus the effects on the spectra from varying these parameters are difficult to disentangle. Applying the GPLVM to a larger number of quasars which span a wide range of parameters will enable a more detailed study of the spectral dependencies with physical quasar properties in the future. 

\section{Current limitations of the model} \label{sec:limitations}

In this work we presented a novel generative model based on a GPLVM, which can handle data sets with heteroscedastic noise and missing data correctly. These two features are crucial for any astrophysical applications with real data. The chosen application in \S~\ref{sec:application} represents a proof of concept and a first demonstration of the model capabilities. However, our model and its application are currently limited by two main factors: First, we only know a limited number of quasars at the moment, for which precise RM black hole mass measurements are available (see \S~\ref{sec:summary} for soon upcoming data from new surveys). The second reason is that the current implementation of the GPLVM is not using state-of-the-art tools, and hence the performance and optimization of our model is very time intensive. 

Multiple future improvements on the implementation of our model are possible and already work in progress, such as for instance the use of a significantly more efficient implementation of Gaussian Proceses, possibly with GPU acceleration, e.g.\ \texttt{gpytorch}\footnote{\url{https://gpytorch.ai/}} \citep{gpytorch} or \texttt{tinyGP}\footnote{\url{https://tinygp.readthedocs.io/en/stable/}}. Additionally, the use of auto-differentiation, such as that implemented in \texttt{jax}\footnote{\url{https://jax.readthedocs.io/en/latest/}}, which automatically differentiates native python code, could also reduce the complexity of the code and might speed it up. 

Nevertheless, the nature of a GPLVM is that it will always scale badly with the number of objects. Naively, the training should take computation time proportional to the cube of the number of training set objects. Furthermore, our adaption of the algorithm to handle data with heteroscedastic uncertainties as well as missing data means that one cannot factorize the matrices in the likelihood functions once, which causes the algorithm to scale badly to large data sets. 

However, it would be possible to use the recent 2D generalization of \texttt{celerite} \citep{DFM2017} to compute the likelihoods, which would then result in a linear scaling with the number of data points \citep[e.g.][]{Gordon2020}. Furthermore, for a better scalability of the GPLVM one could apply mini-batch training, where the training set is split into small batches that are used to update the model coefficients  \citep[e.g.][]{Lalchand2022}. 
While a better implementation will certainly improve the computation time and optimization of our generative model, it will still not scale well to extremely large samples, and thus, the GPLVM is a model for smaller, complex, heterogeneous data sets, where each training point was hard won. 

On the other hand, it might be considered absurd, in the machine-learning or data-driven astrophysics literature, to be training a model with a training set of only 31 objects!
As reverberation-mapping projects proceed, this training set will get larger and the results will (presumably) get better. 
That said, we show that even with this tiny training set, we can predict black hole masses in held-out data about as well as might be expected given the measurement precision. This testifies to the power of this generative model, and the consistency, dimensionality, and information content of quasar spectra. 

Another potential limitation of the model could be the generalisability of the model between the small subset of predominantly low-redshift quasars for which RM measurements exist, and a set of high-redshift quasars for which we ultimately aim to understand their black hole masses. However, upcoming surveys such as SDSS-V \citep{Kollmeier2017} and the Legacy Survey of Space and Time \citep[LSST;][]{LSST2019} on the Rubin Observatory will push the RM measurements to quasars at higher redshifts, with measurements expected for quasars at redshifts up to $z\sim4$. These measurements will help bridge the gap between the lower-redshift quasar population that can be used as a training set for the GPLVM and the higher-redshift quasar population which we ultimately aim to use as a testing set. 

Additionally, since the GPLVM allows us to incorporate missing data or data with large measurement uncertainties in a rigorous way, we could add the spectral features of high-redshift quasars to the training set with a corresponding label for the black hole mass that is either missing completely or has a large measurement uncertainty. However, if one adds significantly more unlabeled than labeled data to the training set, one needs to ensure that the latent space will not only learn the spectral features but also the labels. This could potentially require modifications to the model structure to give sufficient weight to the labels, and will thus be part of future work. 

Finally, a substantial limitation of the model is that the model has been provided essentially no prior knowledge about either the atomic physics or the accretion physics generating the spectrum.
Further improvements could likely be made with models that include a mix of data-driven components with physically motivated components (demonstrated for instance in \citealt{LeistedtHogg2017}).

\section{Summary \& Outlook}\label{sec:summary}

This paper presents a generative model for quasar spectra that generates \textit{simultaneously} both the spectral features of the objects as well as its labels. We chose a GPLVM, which can handle heteroscedastic data sets observed with different telescopes or instruments with measurement uncertainties as well as missing, unobserved or unlabeled data in a principled way. Our model allows to consistently predict quasar properties with limited spectral coverage that could be varying between different objects, as well as with noisy or only partially measured labels in the training set. 

As a first application and proof of concept we apply our model to a data set of $31$ quasars with precise black hole mass measurements obtained via the RM technique and show that the model can predict the black hole mass measurements from the spectral features of an unseen quasar close to the best possible precision. Most importantly, we show that the GPLVM can obtain estimates for the black hole mass of a quasar from a limited spectral region, which has the advantage that no specific emission lines such as \hb\ or \mgii\ that might not always be observable due to atmospheric absorption for instance, are required anymore. 

The scope of this first application is currently still limited for two reasons: First, there are only a very limited number of quasars known which we have precise black hole mass measurements based on the RM technique and second, the current implementation of the model can be significantly improved by using state-of-the-art algorithms as discussed in \S~\ref{sec:limitations}. Both limitations will be overcome in the future: work on a more sophisticated implementation of the algorithm is ongoing, and upcoming surveys such as SDSS-V and LSST promise an increase of $2-3$ orders of magnitude in precise RM measurements for quasars up to redshift $z\sim 4$ within the next decade \citep[e.g.][]{Kollmeier2017, LSST2019}. 
These improvements will enable us to constrain precise black hole masses for quasars at all redshifts, which is crucial for understanding co-evolution of galaxies and their central supermassive black holes across cosmic \citep[e.g.][]{Volonteri2012}. 

\subsection{Future potential applications of the GPLVM} \label{sec:future_applications}
There are numerous possible applications of the here presented generative model, a few of which we will discuss briefly. 

\begin{itemize}
    \item Most single-epoch black hole mass measurements are derived using the width of the \hb\ emission line and scaling relations calibrated to low-redshift quasars with RM measurements \citep[e.g.][]{Grier2017, Park2013, Park2017}. However, as discussed before for quasars at high redshifts of $z\gtrsim 3$ the \hb\ emission line is not observable with ground-based observatories and thus the scaling relations are re-calibrated to the still observable rest-frame UV emission lines, such as \civ\ or \mgii\ \citep[e.g.][]{Coatman2017}. For $z\gtrsim 5$ quasars these emission lines however often fall into regions of significant telluric absorption at near-IR wavelengths, and thus even the single-epoch black hole mass scaling relations cannot be applied to these objects.
    
    Using the GPLVM we can omit all calibration and scaling steps which cause additional uncertainties and possible biases in the black hole mass estimates, since the generative model does not need the full spectral coverage or coverage of a specific emission line to determine the unknown labels. 
    Thus predicting the black hole masses from single-epoch spectra of $z\gtrsim 5$ quasars using generative models such as the GPLVM circumvents these limitations and might therefore result in more accurate predictions than conventional scaling relations.
    
    \item As shown in \S~\ref{sec:results_spectra} the generative model can be used to predict unobserved or ``missing'' spectral regions, which could also be useful for predicting the unabsorbed continuum emission of high-redshift quasars in the Lyman-series forest. For quasars at $z\gtrsim5$ a significant fraction of their continuum emission at wavelengths shorter than Ly$\alpha$ at $\lambda_{\rm rest}=1215.67$~{\AA} is absorbed due to the high fraction of neutral hydrogen in the surrounding intergalactic medium (IGM). However, for analyses of the neutral fraction of the intergalactic medium by means of the IGM damping wing \citep[e.g.][]{Simcoe2012, Davies2018d, Greig2022}, measurements of the IGM opacity in the Ly$\alpha$ or Ly$\beta$ forests \citep[e.g.][]{Fan2006, Becker2015, Eilers2018b, Eilers2019b, Yang2020, Bosman2022}, or measurements of the quasars' proximity zone sizes \citep[e.g.][]{Eilers2017a, Eilers2020, Morey2021, Chen2022}, an accurate knowledge of the quasars' unabsorbed continuum emission is essential. To this end, numerous studies have tried to predict the unabsorbed quasar emission using various approaches, such as PCA \citep[e.g.][]{Suzuki2005, Paris2011, Davies2018, Bosman2021}, neural nets \citep[e.g.][]{Durovcikova2020, LiuBordoloi2021}, or constructing composite spectra of nearest neighbours in low-redshift quasar spectra \citep[e.g.][]{Simcoe2012}, in order to accurately predict the emission. 
    
    However, an important shortcoming of all of these approaches is that they are trained using low-redshift quasars, where there is significantly less absorption from the intervening IGM and the continuum emission can be more easily reconstructed. This approach assumes implicitly that there is no redshift evolution in the spectral shape of quasars. However, we know that this is not a good assumption, since we observe differences in the composite spectra of low- and high-redshift quasars \citep[][]{Shen2019, Yang2021}, e.g. emission lines are often more blueshifted with respect to the quasars' systemic redshifts in high-redshift quasars \citep[e.g.][]{Meyer2019}, which could lead to biases in the continuum reconstruction. The advantage of the GPLVM presented here is that it can deal with data sets with heteroscedastic noise and missing data, and thus we can include the spectra of \textit{both} low- and high-redshift quasars in the training set assuming ``missing'' spectral coverage bluewards of the Ly$\alpha$ line for the high-redshift quasar spectra, which are heavily affected by IGM absorption. 

    \item Several quasar properties require expensive and time consuming observations to be determined, such as for instance measurements of the systemic redshift of a quasar. The most reliable redshift estimates are based on sub-mm emission lines, such as [\ion{C}{2}] at $158\,\mu$m, which are the dominant cooling mechanism of the interstellar medium in the quasars' host galaxies \citep[e.g.][]{CarilliWalter2013}. In contrast, broad rest-frame UV and optical emission lines are subject to strong internal motions or winds in the BLR and thus often displaced from the systemic redshift \citep[e.g.][]{Richards2002, Meyer2019}. However, obtaining sub-mm observations with the Atacama Large Millimetre Array (ALMA) for instance is highly competitive and expensive, and thus one could attempt to infer the quasar's systemic redshifts by means of this generative model, since the velocity shifts of different rest-frame UV and optical lines likely encode information about the quasar's systemic redshift. 
    
\end{itemize}

\acknowledgements
It is our pleasure to thank Joe Hennawi, Robert Simcoe, Hans-Walter Rix, Aaron Barth, Adrian Price-Whelan, Vidhi Lalchand, and Vincent Sitzmann for very helpful discussions. Furthermore, we are grateful to Daesong Park for sharing the HST data, as well as to Joe Hennawi for the use of the computing cluster at UCSB.

This project was developed in part at the 2017 Heidelberg Gaia Sprint, hosted by the Max-Planck-Institut f\"ur Astronomie, Heidelberg.

ACE acknowledges support by NASA through the NASA Hubble Fellowship grant $\#$HF2-51434 awarded by the Space Telescope Science Institute, which is operated by the Association of Universities for Research in Astronomy, Inc., for NASA, under contract NAS5-26555. 

JTS acknowledges funding through the ERC European Research Council (ERC) under the European Union’s Horizon 2020 research and innovation program (grant agreement No 885301).

\software{numpy \citep{numpy}, scipy \citep{scipy}, matplotlib \citep{matplotlib}, astropy \citep{astropy}}

\bibliography{literatur_hz}

\end{document}